\documentclass[prb,twocolumn,showpacs,floats,eqsecnum,amsmath,amssymb,footinbib]{revtex4}
\usepackage[dvips]{graphicx}

\begin{document}



\title{Ferromagnetism in a hard-core boson model}

\author{A. Fledderjohann$^{1}$, A. Langari$^{2}$, E. M\"uller-Hartmann$^{3}$ 
and K.-H. M\"utter$^{1}$}

\affiliation{$^1$Physics Department, University of Wuppertal, 42097 Wuppertal,
Germany}
\affiliation{
$^2$Institute for Advanced Studies in Basic Sciences,
Zanjan 45195-159, Iran}
\affiliation{$^3$ Physics Department, University of Cologne, 50937 K\"oln, Germany}
\date{\today}


\begin{abstract}
\leftskip 2cm
\rightskip 2cm
The problem of ferromagnetism -- associated with a ground state with maximal
total spin -- is discussed in the framework of a hard-core model, which
forbids the occupancy at each site with more than one particle. It is shown
that the emergence of ferromagnetism on finite square lattices crucially
depends on the statistics of the particles. Fermions (electrons) lead to the
well-known instabilities for finite hole densities, whereas for bosons (with
spin) ferromagnetism appears to be stable for all hole densities.

\end{abstract}


\maketitle



\section{Introduction\label{sec1}}

It is generally believed that ferromagnetism should be explained in the
framework of a Hamiltonian for strongly coupled electrons. Such 
a Hamiltonian on a lattice has been proposed 40 years ago by Hubbard.
\cite{hubbard63,gutzwiller63,kanamori63} It contains a nearest neighbour
hopping term $H_t$ and an on-site Coulomb repulsion (of strength $U$) $H_U$
for the electrons.

Both terms are ``blind'' to the spin of the electrons and it is therefore
far from being obvious that such a model should lead to ferromagnetism.
The limit $U\rightarrow\infty$ -- here called ``hard-core condition`` --
enforces that each site can be occupied by at most one electron. In this
limit, the number $Q$ of electrons which can be accomodated on a lattice
of $N$ sites is limited $Q\leq N$. In the charge sector $Q=N$ every spin
configuration of the electrons is a possible ground state. On a square
lattice this degeneracy is lifted in the sector $Q=N-1$ with one hole.
Nagaoka\cite{nagaoka66} constructed the ground state and found that the
total spin $S(Q)$ of the electrons is maximal:
\begin{eqnarray}
S(Q) & = & \frac{Q}{2},\,\,\,Q=N-1
\end{eqnarray}
a clear signature for ferromagnetism.

There has been an intensive search for ferromagnetism in the $t-J$ model
with $N_h=N-Q$ holes. Numerical results on finite clusters \cite{riera89,
fehske91} at vanishing exchange coupling 
($\alpha=0$, see Eq.(\ref{H_tJ})) show that the ground
state does not have maximal spin except for the Nagaoka case $Q=N-1$.
The instability of the Nagaoka ferromagnetic state in sectors with 2 and
more holes has been investigated with variational ground states of the
Gutzwiller type.\cite{shastry90,basile90,vonderlinden91,hanisch93}
It was found that there exists a critical hole density $\delta=N_h/N=1-\rho$
such that the ground states cannot be ferromagnetic for $\delta>\delta_{crit.}$.
Improvements of the variational ansatz led to a successive decrease of
$\delta_{crit.}$ for the $2D$ case
\begin{eqnarray}
\delta_{crit.} & = & 0.49\cite{shastry90},\,\,0.41\cite{basile90},
\,\,0.29\cite{vonderlinden91,hanisch93},\,\,0.251\cite{wurth96}\,. \nonumber
\end{eqnarray}
On the other hand Barbieri, Riera and Young\cite{barbieri90} have argued
that the ferromagnetic ground state emerges in the thermodynamical limit
$N\rightarrow\infty$, if the hole density 
vanishes ($\delta\rightarrow 0$).
The existence of a lower bound $\delta>\delta_{crit.}$ for the hole
density $\delta$, where ferromagnetism is not possible, is indeed a quite
general feature, which has been found also in higher dimensions and for
various types of lattices.\cite{hanisch95}
An introduction to ferromagnetism in the Hubbard model can be found in
Ref.[\onlinecite{tasaki03}].

In this situation one might ask for the reason why hard-core models with
fermions (i.e. electrons) fail to ``explain'' ferromagnetism on finite
clusters (except for the 1 hole case). We want to demonstrate in this
paper that this failure
can be traced back to the anticommutation relations for the fermion
operators. For this purpose, we  substitute in the $t-J$ model the
fermionic degrees of freedom by bosonic ones. In order to facilitate the
comparison of the fermionic and bosonic version, we assume that the
bosons carry here as well spin 1/2 and experience the same antiferromagnetic 
interaction.\cite{footnote1}

Therefore the Hamiltonians for a hard-core model with bosons ($b$) and
fermions ($f$) are both blind with respect to the spin. Nevertheless,
we find pronounced differences (and similarities) in the ground state
energies $E_i(Q,S_i(Q))$, $i=f,b$ and ground state spins $S_i(Q)$,
$i=f,b$.

(a) For $Q=0,1,N-1$ the ground state energies and ground state spins
coincide
\begin{eqnarray}
E_b(Q,S_b(Q)) & = & E_f(Q,S_f(Q))\label{E_bf}\\
S_b(Q) & = & S_f(Q)=Q/2\label{S_bf}
\end{eqnarray}
This means in particular that the construction of the Nagaoka ferromagnet
state holds for fermions and for bosons. The restriction to the one hole
sector $N_h=1$, $Q=N-1$ is crucial. Moreover one observes a hole-particle
symmetry for $Q=1$,
\begin{eqnarray}
E_i\big(Q,S_i(Q)\big) & = & E_i\big(N-Q,S_i(N-Q)\big)\quad i=f,b\,.
\nonumber\\ \label{p-h-symm}
\end{eqnarray}

(b) For $Q=2$ the ground state energies coincide [(\ref{E_bf}) for $Q=2$],
but the ground state spins are different
\begin{eqnarray}
S_f(Q=2)=0 & \quad & S_b(Q=2)=1\,,\label{remark_Q2}
\end{eqnarray}
i.e. in the bosonic case the ground state spin is maximal.

(c) For $Q=3,\ldots,N-2$ ground state energies and ground state spins
are different for the fermionic and bosonic version. In the bosonic version
the ground state spin $S_b(Q)=Q/2$ is maximal as shown in Appendix \ref{appendix_a}.
The ground state energies $E_b(Q,S_b(Q))$ show the hole--particle
symmetry (\ref{p-h-symm}) for these $Q$-values. This symmetry is not present
in the fermionic version. In the latter  case the ground state spin
$S_f(Q)$ is ``erratic''.\cite{riera89}

We think that these properties are interesting enough to justify a
detailed study of the ground state properties of the bosonic $t-J$ model
and to compare its properties with the corresponding properties in the
fermionic version.



The outline of the paper is the following:

In Section\ref{sec2} we review the definition of the $t-J$ model
and point out the differences between the fermionic and bosonic
version. Consequences for the ground state energies on the smallest
($2\times 2$) cluster are discussed in Section\ref{sec3}.
The phase diagram of the bosonic $t-J$ model in dimensions
$D=2$ and $D=1$ is discussed
in Section\ref{sec4} and \ref{sec5}, respectively.





\section{The bosonic version of the $t-J$ model in two dimensions\label{sec2}}

Let us first recall the definition of the $t-J$ Hamiltonian with
fermionic degrees of freedom:
\begin{eqnarray}
H & = & t\Big(H_t+\alpha H_J\Big)\quad\alpha=J/t\,.\label{H_tJ}
\end{eqnarray}
The hopping term
\begin{eqnarray}
H_t & = & -{\cal P}\sum_{\langle x,y\rangle}\sum_{\sigma}\left(
c_{\sigma}^+(x)c_{\sigma}(y) + \mbox{h.c.} \right){\cal P}\label{H_t}
\end{eqnarray}
and the spin exchange part
\begin{eqnarray}
H_J & = & {\cal P}\sum_{\langle x,y\rangle}\Big({\bf S}(x){\bf S}(y)
-\frac{1}{4}n(x)n(y)\Big){\cal P}\label{H_J}
\end{eqnarray}
can be expressed in terms of creation ($c^+_{\sigma}(x)$) and annihilation
($c_{\sigma}(x)$) operators for the electrons, which obey anticommutation
relations
\begin{eqnarray}
\big\{c^+_{\sigma}(x),c^+_{\sigma'}(x')\big\} & = & \big\{c_{\sigma}(x),c_{\sigma'}
(x')\big\}
=0\nonumber\\
\big\{c^+_{\sigma}(x),c_{\sigma'}(x')\big\} & = & \delta_{\sigma,\sigma'}\delta_{x,x'};
\nonumber\\ \label{anticommutation}
\end{eqnarray}
\begin{eqnarray}
n_{\sigma}(x) & = & c^+_{\sigma}(x)c_{\sigma}(x)
\end{eqnarray}
is the number operator for an electron at site $x$ with spin $\sigma$.
Owing to the ``hard-core condition'' which is imposed by the projector
${\cal P}$ double occupancy is forbidden i.e.
$\sum_{\sigma}n_{\sigma}(x)=0,1$. The latter can be derived from a Hubbard
model\cite{hubbard63} with infinite on-site Coulomb repulsion ($U/t\rightarrow\infty$).

The construction of a state with $Q$ electrons
\begin{eqnarray}
|x_1^{\sigma_1},x_2^{\sigma_2},\ldots,x_Q^{\sigma_Q},\rangle & = &
c_{\sigma_1}^+(x_1)\ldots c_{\sigma_Q}^+(x_Q) |0\rangle
\end{eqnarray}
by application of creation operators to the vacuum $|0\rangle$ demands
an ordering of all sites on the 2 dimensional square lattice. Owing
to the anticommutation rules (\ref{anticommutation}) a different
ordering (e.g. with $x_1^{\sigma_1},x_2^{\sigma_2}$ interchanged)
leads to a state which might differ from the former one in sign. In
the following we will denote the traditional model (\ref{H_tJ})--(\ref{H_J})
with fermions as ``fermionic $t-J$ model''.

Let us now turn to the ``bosonic $t-J$ model'', which we simply define
by substituting the anticommuting creation and annihilation operators

\begin{eqnarray}
c_{\sigma}^+(x)\rightarrow a_{\sigma}^+(x) &  \quad &
c_{\sigma}(x)\rightarrow a_{\sigma}(x)
\end{eqnarray}
by commuting ones:
\begin{eqnarray}
\big[a^+_{\sigma}(x),a^+_{\sigma'}(x')\big] & = &
\big[a_{\sigma}(x),a_{\sigma'}(x')\big]=0
\nonumber\\
\big[a^+_{\sigma}(x),a_{\sigma'}(x')\big] & = &
\delta_{\sigma,\sigma'}\delta_{x,x'}\,.
\nonumber\\ \label{commutation}
\end{eqnarray}
The eigenvalues of the number operator
\begin{eqnarray}
n_{\sigma}(x) & = & a^+_{\sigma}(x)a_{\sigma}(x)
\end{eqnarray}
are restricted again to $n_{\sigma}(x)=0,1$ due to the ``hard-core condition''.
This model has been discussed first in references (\onlinecite{bonisegni01,
bonisegni02,altman03,smakov04}).

Quite recently a generalized Hubbard model with fermionic
and bosonic degrees of freedom has been proposed and investigated
in order to study a mixture of ultracold bosonic
and fermionic atoms in an optical lattice.\cite{lewenstein04}

The construction of a state with $Q$ bosons by application of creation
operators on the vacuum, however, is symmetric under the permutation
of sites -- due to the commutation relations (\ref{commutation}). There
does not exist a ``sign problem''. This is the only difference between
the fermionic and bosonic version of the $t-J$ model [cf. Ref. (16)].
One can easily verify, that the action of the spin exchange term (\ref{H_J}) is the
same in the fermionic and bosonic version.
On the other hand the hopping term $H_t$ acts indeed in a different way
on the fermionic and bosonic states. In the latter case $H_t$ can be
expressed in terms of nearest neighbour permutations $P(x,y)$ which
interchange a particle and a hole at sites $x$ and $y$:
\begin{eqnarray}
H_t & = & -\sum_{\langle x,y\rangle}P(x,y)\Big(n(x)n_h(y)+n(y)n_h(x)\Big)\,.
\nonumber\\ \label{H_tb}
\end{eqnarray}
Here $n(x)=n_+(x)+n_-(x)=0,1$ is the number of bosons at site $x$.

\begin{figure}[ht!]
\centerline{\hspace{0.0cm}\includegraphics[width=9.0cm,angle=0]{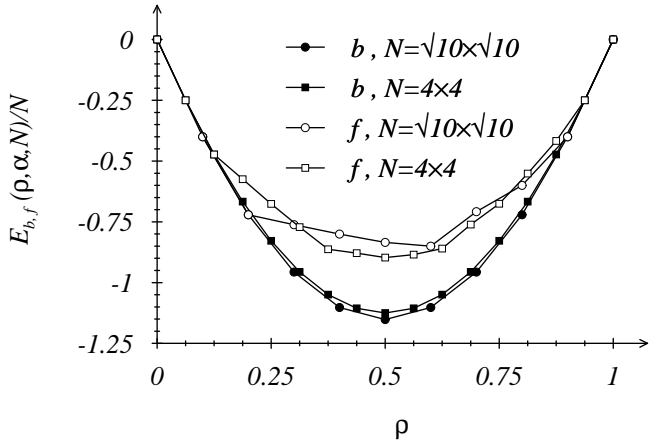}
\hspace{0.0cm}}
\caption{Ground state energies per site of a periodic $4\times 4$ and a
$\sqrt{10}\times\sqrt{10}$ lattice for both fermionic and bosonic $t-J$ models
at $\alpha=0$.}
\label{fig1}
\end{figure}

In Appendix \ref{appendix_a}
the Hamiltonian (\ref{H_tb}) is proven to have a ferromagnetic ground state
for all $Q$. Moreover it is shown that the ground state energies of the bosonic 
$t-J$ model at $\alpha=0$ are symmetric under the particle--hole transformation
\begin{eqnarray}
E_b(Q,\alpha=0,N) & = & E_b(N-Q,\alpha=0,N)\label{E_b_ph}
\end{eqnarray}
as can be seen in Fig. \ref{fig1}.

Here we also show the ground state energy for hard-core fermions, where the
hole--particle symmetry (\ref{E_b_ph}) is lost (except for the
Nagaoka case $Q=N-1$)
\begin{eqnarray}
E_f(Q,\alpha=0,N) & \ne & E_f(N-Q,\alpha=0,N)\,.
\end{eqnarray}
Moreover the ground state energies of the hard-core bosons are below
those of the hard-core fermions
\begin{eqnarray}
E_b(Q,\alpha=0,N) & \leq & E_f(Q,\alpha=0,N)\,..
\end{eqnarray}
Comparing the results from lattices $L\times L$, $L=\sqrt{10},4$ we find
that the finite-size dependence is rather small and smooth in the bosonic
case, in contrast to the fermionic case. The result presented in Fig. \ref{fig1}
were obtained with a Lanczos algorithm. Concerning the fermionic version,
they agree with Ref. (\onlinecite{riera89}) (Riera, Young).\cite{footnote2}

In both versions (fermionic and bosonic) the $t-J$ Hamiltonian (\ref{H_tJ})
conserves the total charge
\begin{eqnarray}
Q & = & \sum_x\big(n_+(x)+n_-(x)\big)
\end{eqnarray}
and total spin
\begin{eqnarray}
{\bf S} & = & \frac{1}{2}\sum_x{\bf\sigma(x)}\label{s_total}
\end{eqnarray}
such that the eigenvalues of ${\bf S}^2=S(S+1)$ can be used to characterize
the eigenstates of the Hamiltonian.
Ferromagnetic eigenstates $|F,Q\rangle$ have maximal spin $S=Q/2$. These
states are simultaneous eigenstates of the hopping part $H_t$ and the
spin coupling part $H_J$
\begin{eqnarray}
tH_t|F,Q\rangle & = & E_F(Q)|F,Q\rangle\label{H_t_F}\\
H_J|F,Q\rangle & = & 0\label{H_J_F}
\end{eqnarray}
such that the eigenvalues $E_F(Q)$ for the $t-J$ Hamiltonian (\ref{H_tJ})
are independent of $\alpha$. This is a consequence of the fact that the
hopping part $H_t$ is blind with respect to the spin of the electrons.
There is no difference in the hopping of spin-up and spin-down particles.
On the other hand, the ferromagnetic eigenstates $|F,Q\rangle$ are
totally symmetric under any permutation of spin-up and spin-down particles.
Therefore, the application of the nearest neighbour spin exchange coupling
\begin{eqnarray}
h_J(x,y) & = & \big(P(x,y)-1\big)n(x)n(y)\label{SU3_breaking}
\end{eqnarray}
yields zero:
\begin{eqnarray}
h_J(x,y)|F,Q\rangle & = & 0\,.
\label{h_J_xy_FQ}
\end{eqnarray}

\section{Ground state properties of the fermionic and bosonic $t-J$
model on a $2\times 2$ plaquette\label{sec3}}

On a plaquette with 4 sites ($2\times 2$) the $t-J$ Hamiltonian can
be diagonalized analytically.\cite{fl04} It is interesting to compare the
ground state energies $E^{(p)}(Q)$ and total spins $S^{(p)}(Q)$
in the sectors with $Q$ spin-1/2 particles for the fermionic and
bosonic version. It turns out

\vspace{0.2cm}

\hspace{-0.3cm}{\bf (1)} they are the same in the sectors $Q=0,1,3,4$

\begin{itemize}
\item[]
\begin{eqnarray}
\hspace{-0.0cm}E^{(p)}(0) & = & 0\hspace{4.0cm} S^{(p)}=0\hspace{0.8cm}\\
E^{(p)}_F(1) & = & -2t\,\,\,(\alpha\geq 0)\hspace{2.3cm} S^{(p)}=\frac{1}{2}\\
E^{(p)}(3) & = & \left\{\begin{array}{ll}-2t\hspace{0.0cm} & (F)\,\,\,S_F^{(p)}=\frac{3}{2}\\
-t\Big(\alpha+\sqrt{\frac{\alpha^2}{4}+3}\Big)\hspace{0.0cm} & (B)\,\,\,S^{(p)}=\frac{1}{2}\\
-t\Big(\frac{3\alpha}{2}+1\Big) & (C)\,\,\,S^{(p)}=\frac{1}{2}
\end{array}\right.\label{E_p_3}\\
E^{(p)}(4) & = & -3t\alpha\,\,\,(\alpha\geq 0)\hspace{2.12cm} S^{(p)}=0
\end{eqnarray}
where the $\alpha$ intervals $(F)$, $(B)$, $(C)$ are defined by
\begin{eqnarray}
(F): & \hspace{0.5cm} & 0\leq\alpha\leq\alpha_F\Big(\frac{3}{4}\Big)=0.262 \nonumber\\
(B): & & \alpha_F\Big(\frac{3}{4}\Big)\leq\alpha\leq 2 \nonumber\\
(C): & & \alpha\geq 2 \nonumber
\end{eqnarray}

Ground state energies with index ``$F$'' correspond to ground states with
maximal spin $S=Q/2$. In the sector with three particles ($Q=3$)
one finds the same energy in the
fermionic and bosonic version of the $t-J$ model for all $\alpha\geq 0$.
\end{itemize}

\vspace{0.2cm}

\hspace{-0.3cm}{\bf (2)} they are different in the sector $Q=2$
\begin{itemize}
\item[]

\vspace{0.2cm}

a) In the fermionic version,
\begin{eqnarray}
E^{(p)}(2) & = & -\frac{t}{2}\Big(\alpha+\sqrt{\alpha^2+32}\Big)
\nonumber
\\ & & \hspace{0.3cm}
\alpha> 0,\hspace{0.3cm}S^{(p)}(2)=0\nonumber\\\label{E_p_Q2_f}
\end{eqnarray}
the ground state is nondegenerate and has total spin 0.

b) In the bosonic version:
\begin{eqnarray}
\hspace{0.8cm}E^{(p)}_F(2) & = & -2t\sqrt{2}\hspace{2.6cm}\hspace{0.3cm}\nonumber\\
 & & 0\leq\alpha\leq\alpha_F\Big(\frac{1}{2}\Big)=\sqrt{2},
\hspace{0.3cm}S^{(p)}_F(2)=1\nonumber\\
 & & \label{E_p_Q2_b1}\\
E^{(p)}(2) & = & -\frac{t}{2}\Big(\alpha+\sqrt{\alpha^2+16}\Big)\hspace{0.5cm}
\hspace{0.3cm}\nonumber\\
 & & \alpha> \alpha_F\Big(\frac{1}{2}\Big),\hspace{0.3cm}
S^{(p)}(2)=0\nonumber\\
 & & \label{E_p_Q2_b2}
\end{eqnarray}
we find first a ferromagnetic ground state with maximal spin 1 and for larger
$\alpha$ values degenerate spin 0 ground states.
\end{itemize}

Note that the ground state energy of the fermionic model (\ref{E_p_Q2_f})
is below that of the bosonic model (\ref{E_p_Q2_b1}), (\ref{E_p_Q2_b2})
for $\alpha>0$. The two energies coincide at $\alpha=0$.

\begin{figure}[ht!]
\centerline{\hspace{0.0cm}\includegraphics[width=8.0cm,angle=0]{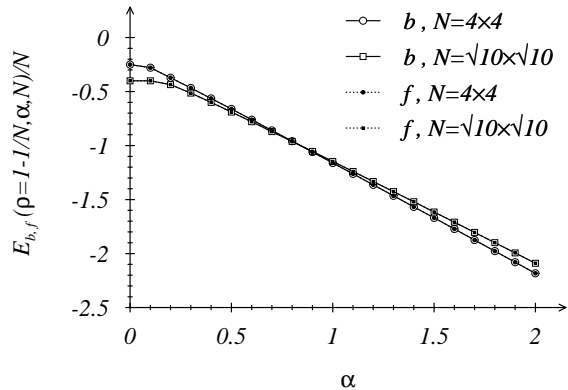}
\hspace{0.0cm}}
\caption{$\alpha$-dependence of ground state energies in the 1 hole sector
of a $L\times L$ lattice for $\alpha>0$. Bosonic and fermionic
$t-J$ ground state energies for $L=\sqrt{10},4$ coincide.}
\label{fig2}
\end{figure}

It is remarkable to note, that some of these similarities can be found on
all square lattices $L\times L$, $L=2,\sqrt{10},4$.
Most important is the equality
\begin{eqnarray}
E_f(Q=N-1,\alpha,N) & = & E_b(Q=N-1,\alpha,N)\nonumber\\
 & & \quad N=L^2,\,\,\alpha>0
\end{eqnarray}
in the 1 hole sector for all $\alpha>0$. The $\alpha$ dependence is
shown in Fig. \ref{fig2}.

In the sector with 2 particles ($Q=2$) at $\alpha=0$ the ground state
energies are the same, but the ground state spin is always $S_f(Q)=0$
in the fermionic, but $S_b(Q)=1$ in the bosonic version, as already
stated in (\ref{remark_Q2}).

\section{The phase diagram of the bosonic $t-J$ model in
two dimensions\label{sec4}}

In this paper we are mainly concerned with the question, whether the
ground state with $Q$ (spin-1/2) particles has maximal total spin
$S=Q/2$. The answer to this question depends on the strength $\alpha$
of the spin exchange part (\ref{H_tJ}) which prefers (for $\alpha>0$)
antiferromagnetic ordering with total spin $S=0$.

Owing to the property (\ref{H_J_F}) the expectation value
of the spin exchange part
\begin{eqnarray}
\varepsilon_J(\rho,\alpha) & = & \frac{1}{N}\langle E(Q,\alpha)|H_J|E(Q,\alpha)\rangle
\label{eps_J}
\end{eqnarray}
can be considered as an order parameter; it vanishes if the ground
state $|E(Q,\alpha)\rangle$ in the sector with $Q$ particles has maximal
spin $S=Q/2$, but
is nonzero in all other cases.

We will demonstrate for the bosonic version of the $t-J$ model on
$L\times L$ clusters that the ferromagnetic regime:
\begin{eqnarray}
\varepsilon_J(\rho,\alpha) & = & 0\quad\mbox{for\,\,}0\leq\alpha\leq
\alpha_F(\rho=Q/N)\label{eps_J_0}
\end{eqnarray}
is bounded by a curve $\alpha=\alpha_F(\rho=Q/N)$ depending on
the hole density $\delta=1-\rho$.

\begin{figure}[ht!]
\centerline{\hspace{0.0cm}\includegraphics[width=11.0cm,angle=-90]{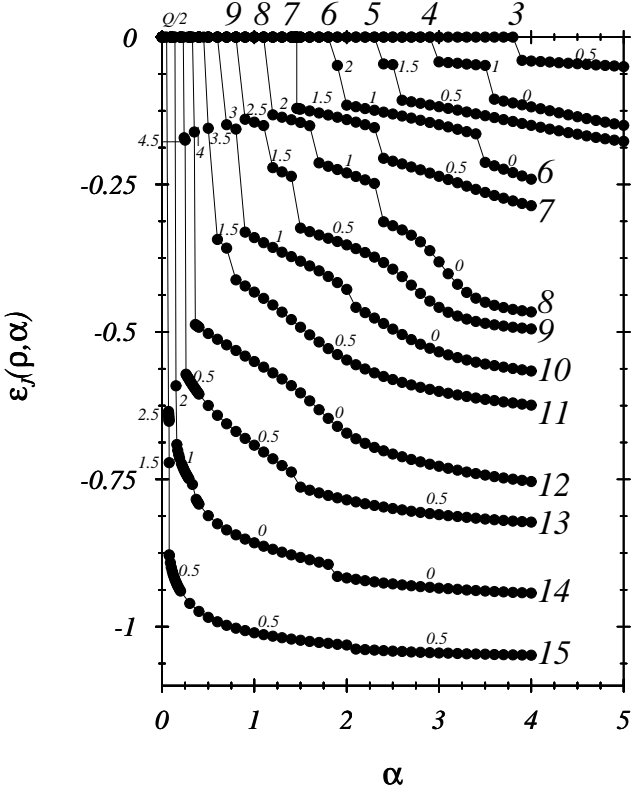}
\hspace{0.0cm}}
\caption{Order parameter $\varepsilon_J(\rho,\alpha)$
for a $N=4\times 4$ cluster with charge sectors $Q=3,...,15$.
The order parameter for each $Q$ is turning away from zero
at $\alpha=\alpha_F(\rho=Q/N)$. Small numbers denote the
total spin of the bosonic $t-J$ ground states.}
\label{fig3}
\end{figure}

In our numerical calculations of the ground states $|E(Q,\alpha)\rangle$,
$Q=N-1,N-2,...$ in the bosonic $t-J$ model with periodic boundary conditions,
we only found ferromagnetic states $|F\rangle=|E_F(Q)\rangle$ for $\alpha$
small enough.
Fig. \ref{fig3} shows the order parameter $\varepsilon_J(\rho,\alpha)$ on
a $4\times 4$ cluster in the charge sectors $Q=3,4,..,15$.

\begin{figure}[ht!]
\centerline{\hspace{0.0cm}\includegraphics[width=8.0cm,angle=-90]{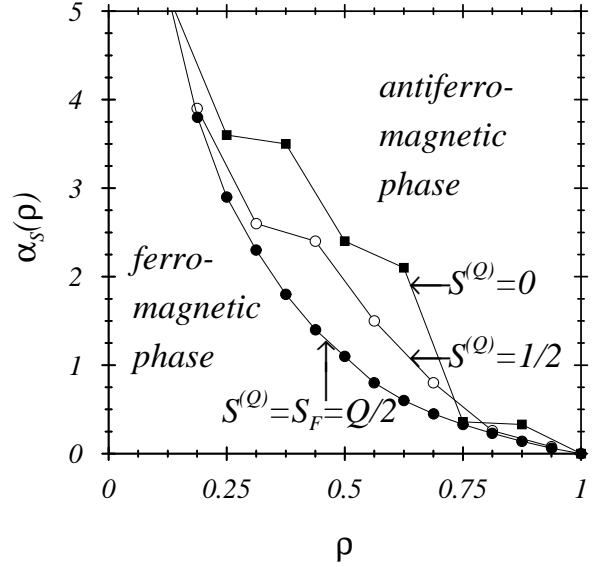}
\hspace{0.0cm}}
\caption{$\rho$-dependence of the lines $\alpha_0(\rho)$,
$\alpha_{1/2}(\rho)$, $\alpha_F(\rho)$ for a $N=4\times 4$ cluster.}
\label{fig4}
\end{figure}

The phase
boundary $\alpha_F(\rho=Q/16)$ can be read off from those $\alpha$-values
where $\varepsilon_J(\rho,\alpha)$ changes first from 0 to a
nonvanishing value. Obviously, $\alpha_F(\rho)$ is monotonically decreasing
with $\rho$ and vanishes for $\rho=1$. Its $\rho$-dependence is shown in
Fig. \ref{fig4}.

For increasing values of $Q$, we observe more and more jumps
in the order parameter $\varepsilon_J(\rho,\alpha)$. E.g. for
$Q=8$ we find three of these jumps. Each of them signals the
transition to a new ground state with increasing values of
$\alpha$. The ground states differ in their total spin $S$
which is denoted by the integer ($Q$ even) and halfinteger
($Q$ odd) numbers at the branches between two jumps. E.g.
for $Q=8$:

\begin{eqnarray}
S=4 & \mbox{\quad for\quad} & 0.0\leq\alpha\leq 1.1=\alpha_F(1/2)
\hspace{0.5cm}|E_{F}\rangle\nonumber\\
S=2 & \mbox{\quad for\quad} & 1.2\leq\alpha\leq 1.6
\hspace{2.25cm}|E_{B}\rangle\nonumber\\
S=1 & \mbox{\quad for\quad} & 1.7\leq\alpha< 2.4=\alpha_0(1/2)
\hspace{0.60cm}|E_{C}\rangle\nonumber\\
S=0 & \mbox{\quad for\quad} & \hspace{1.0cm}\alpha\geq 2.4
\hspace{2.15cm}|E_{D}\rangle\nonumber
\end{eqnarray}

So far we were not able to localize the spin couplings for
the ground state with total spin $S=3$, which we expect to emerge
in the interval $1.1<\alpha<1.2$.
For increasing values of $\rho=Q/N$, $\alpha_F(\rho)$ approaches
zero and it is more and more difficult to localize the $\alpha$ values
for those ground states with higher spin $S=Q/2-1,Q/2-2,..$. We
performed higher resoluted searches with step width $\Delta\alpha=0.001$
and found e.g. in the $Q=15$ sector the ground state with total spin
$S=1.5$ at $\alpha=0.077$. In contrast, a similar search for the
$S=2.5$ ground state in the $Q=7$ sector failed.

As an alternative we also tried to localize those $\alpha$ values:
\begin{eqnarray}
\alpha_0(\rho),\,\,\rho=\frac{Q}{N}, & & Q\mbox{\,\, even},\,\, S=0\nonumber\\
\alpha_{\frac{1}{2}}(\rho),\,\,\rho=\frac{Q}{N}, & & Q\mbox{\,\, odd\,},\,\,
S=\frac{1}{2}\nonumber
\end{eqnarray}
where the total spin $S$ of the ground state is first minimal.
The curves $\alpha_0(\rho)$, $\alpha_{\frac{1}{2}}(\rho)$, and
$\alpha_F(\rho)$ are shown in Fig. \ref{fig4}.

\begin{figure}[ht!]
\centerline{\hspace{0.0cm}\includegraphics[width=9.0cm,angle=0]{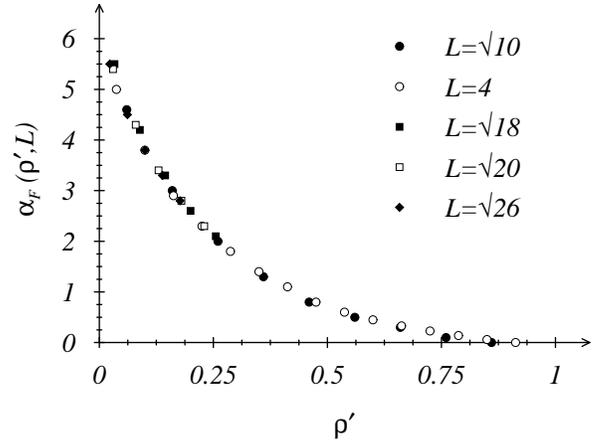}
\hspace{0.5cm}}
\caption{Finite-size effects of $\alpha_F(\rho)$
by comparison of results on $L\times L$ lattices for $L=\sqrt{10},4,
\sqrt{18},\sqrt{20},\sqrt{26}$.
The finite-size effects has been accounted for by 
$\rho'=\rho-\frac{1.4}{N}$.
}
\label{fig5}
\end{figure}

All curves are monotonically decreasing with $\rho$.
We have looked for the finite-size dependence of the curve $\alpha_F(\rho)$
by comparing results on $L\times L$ lattices for
\begin{eqnarray}
L & = & \sqrt{10},4,\sqrt{18},\sqrt{20},\sqrt{26}\,.\nonumber
\end{eqnarray}
On the larger systems with $L>4$ we assumed that the ground state
spin $S_b=Q/2$ is maximal and the curve $\alpha_F(\rho)$ was extracted
from those $\alpha$-values, where the spin is lowered first by one
or two units. As can be seen from 
Fig. \ref{fig5},
the finite-size effects
can be rather well accounted for by an effective charge density
\begin{eqnarray}
\rho' & = & \rho-\frac{c}{N},\quad c=1.4\nonumber
\end{eqnarray}
such that $\alpha_F(\rho')$ scales with the system size $N=L\times L$.

A similar finite-size analysis for the curves $\alpha_0(\rho)$ and 
$\alpha_{1/2}(\rho)$, which mark the boundaries of the antiferromagnetic
phase with minimal ground state spin demands simulations on much larger
systems. They might be accessible with Monte Carlo techniques if it is
possible to determine the ground state spin with high accuracy. The
question of physical interest is, wether the transition region between
the ferro- and antiferromagnetic phase in Fig. \ref{fig4} shrinks to
zero in the thermodynamical limit.

\section{Comparison of the bosonic and fermionic $t-J$ model in one dimension
\label{sec5}}

\begin{figure}[ht!]
\centerline{\hspace{0.0cm}\includegraphics[width=9.0cm,angle=0]{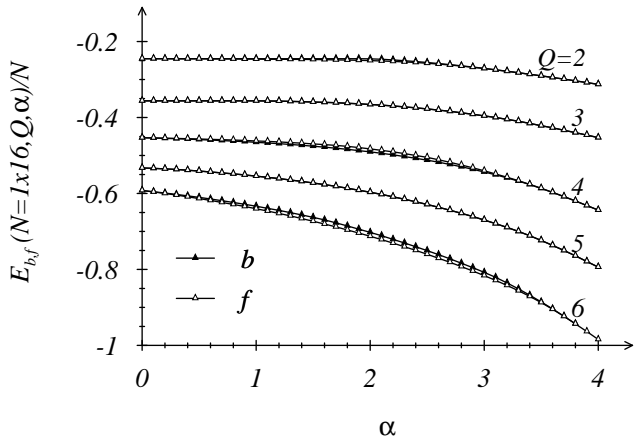}
\hspace{0.0cm}}
\caption{Comparison of the bosonic and fermionic ground state energies
per site ($E_{b,f}(N,Q,\alpha)/N$)
for a one dimensional periodic chain $N=16$ ($D=1$)
for charges $Q=2,3,4,5,6$.}
\label{fig6}
\end{figure}

In one dimension the difference between the bosonic and fermionic version
of the $t-J$ model can be absorbed in the boundary conditions. It turns
out that the ground state energies with periodic boundary conditions
on a $N=16$ site system coincide for all $\alpha$ if $Q$ is odd
\begin{eqnarray}
E_b\big(Q,S_i(Q),\alpha,N\big) & = & E_f\big(Q,S_i(Q),\alpha,N\big)\,.
\label{Eb_Ef_al0}
\end{eqnarray}
For $Q$ even one observes slight differences for $\alpha\simeq 2$,
as can be seen from Fig. \ref{fig6}.

At $\alpha=0$ the equality (\ref{Eb_Ef_al0}) strictly holds for all
$Q$-values and all the ground state energies satisfy the hole--particle
symmetry (\ref{p-h-symm}).

Concerning the total spin of the ground state $S_i(Q)$ $i=b,f$ at
$\alpha=0$ we observe
\begin{itemize}
\item[a)]
a maximal value $S_b(Q)=Q/2$ for all $Q$

\item[b)]
a maximal value $S_f(Q)=Q/2$ only for odd $Q$

\end{itemize}
For $Q$ even the ground states appear to be degenerate
with different values for $S_f(Q)$ of the total spin. This leads to
the ``erratic'' behaviour observed already in the two dimensional
system $D=2$.

\begin{figure}[ht!]
\centerline{\hspace{0.0cm}\includegraphics[width=9.0cm,angle=0]{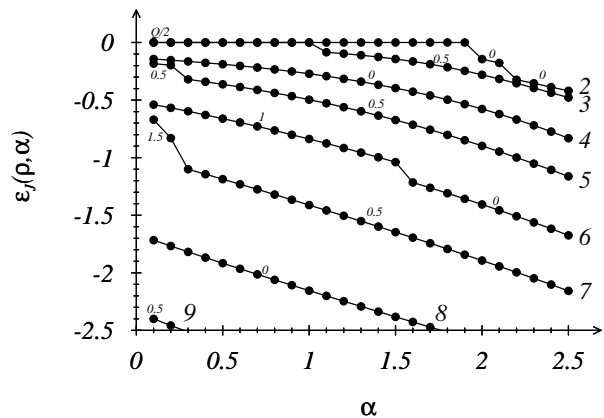}
\hspace{0.0cm}}
\caption{Order parameter $\varepsilon_J(\rho,\alpha)$
for a $N=1\times 16$ cluster with charge sectors $Q=3,...,15$.
Small numbers denote the
total spin of the bosonic $t-J$ ground states.}
\label{fig7}
\end{figure}

Differences between the bosonic $t-J$ model for
$D=1$ and $D=2$ become apparent in the variation of the order parameter
(\ref{eps_J}) and the total spin $S_b(Q)$ with $\alpha$. They are shown
in Fig. \ref{fig7} for $D=1$ and in Fig. \ref{fig3} for $D=2$.

The main difference between the two cases $D=2$ and $D=1$ is the following:

The ferromagnetic domain $0\leq\alpha\leq\alpha_F(\rho)$ -- where the
total spin is maximal -- shrinks to zero in the 1$D$ case. In particular
for $Q>3$, the boundary $\alpha_F(\rho)$ turns out to be below 
$0.1$ already on finite chains.

\section{Discussion and perspectives\label{sec6}}

In this paper we have demonstrated that the emergence of ferromagnetism
in a hard-core model crucially depends on the statistics of the constituent
particles, i.e. whether they are fermions or bosons. We studied the $t-J$
model in the (traditional) fermionic ($f$) and in a bosonic ($b$) version
and compared ground state energies $E_i(Q,S_i(Q),\alpha)$ and total spins
$S_i(Q,\alpha)$, $i=f,b$, in both models. In particular, we found for
$\alpha=0$ that -- in contrast to hard-core fermions -- hard-core bosons
(with spin 1/2) have a ferromagnetic ground state with maximal spin
$S_b(Q)=Q/2$ in all charge sectors Q.

This means that the corresponding Hamiltonian
$H_t$ (\ref{H_tb}) leads intrinsically to a ferromagnetic interaction
between spin-up and spin-down particles, as we have demonstrated  first
on finite square lattices $L\times L$, $L=2,\sqrt{10},4$ numerically.
A general proof is given in Appendix \ref{appendix_a}.

The ground state energies for hard-core bosons and fermions behave
differently on finite lattices. They show hole--particle symmetry
(\ref{E_b_ph}) in the bosonic version, whereas this
symmetry is not present in the fermionic version. We do not know whether
this breaking is a finite-size effect and whether the ground state
energies (Fig. \ref{fig1}) converge to each other in the thermodynamical
limit. If we switch on the spin exchange coupling $\alpha$ [(\ref{H_J}),
(\ref{SU3_breaking})], the bosonic $t-J$ model in two dimensions
($D=2$) develops a ferromagnetic regime $0\leq\alpha\leq\alpha_F(\rho)$,
where $\alpha_F(\rho)$ is monotonically decreasing with $\rho$. For
$D=1$, $\alpha_F(\rho)$ shrinks to zero.


Ferromagnetism is observed in the fermionic and bosonic $t-J$ model in
the sectors $Q=1$ (1 particle) and $Q=N-1$ (1 hole). Here, the two $t-J$
models are unitary equivalent, which implies that the sign problem for
the electrons can be absorbed in a unitary transformation.\cite{tasaki89}
In other words ``electrons behave as bosons''.
This behaviour can be found as well for hard-core electrons in $1D$ with
periodic boundary conditions and $Q$ odd.

On one hand all ground state energies $E_f(Q,\alpha=0)$ $Q=1,..,N-1$
show the hole--particle symmetry (\ref{p-h-symm}) and are strictly identical
with the corresponding quantities $E_b(Q,\alpha=0)$ in the bosonic
version. The ground state spins $S_f(Q)$ are maximal for $Q$ odd,
but this is not the case for $Q$ even, where the ground state appears
to be degenerate with different total spins.


Let us finally comment on
possible more realistic hard-core boson models for ferromagnetism.
On one hand these bosons have to carry spin, on the other hand the
spin statistics theorem demands integer spin for bosons.
It is quite straightforward to extend the hopping Hamiltonian
(\ref{H_tb}) to spin 1 bosons. In this case, the number of particles
at site $x$:
\begin{eqnarray}
n(x) & = & n_1(x)+n_{-1}(x)+n_0(x)=0,1
\end{eqnarray}
has to be identified with the sum of spin $1,-1,0$ particles. Such
a Hamiltonian is again invariant under particle--hole transformations.
It defines a hard-core boson model with 4 degrees
of freedom at each site: 3 for the spin-1 particles, 1 for the holes.






\begin{appendix}

\section{The ground state of the hopping Hamiltonian (\ref{H_tb})
\label{appendix_a}}

We want to present first a general proof that the ferromagnetic state
with all spins up ($N_+=Q$) is indeed a ground state of $H_t$ (\ref{H_tb})
in the sector with $Q$ particles. Here, the permutation operator
$P(x,y)$ for a spin ($+$) and a hole can be represented:
\begin{eqnarray}
P(x,y) & =  & \frac{1}{2}\Big(1+\vec\tau(x)\vec\tau(y)\Big)
\end{eqnarray}
in terms of Pauli matrices at neighbouring sites $x$ and $y$ which
act on particle $\chi_+$ and hole $\chi_h$ states as
\begin{eqnarray}
\tau_3(x)\chi_{j}(x) & = & \big(2n_+(x)-1\big)\chi_{j}(x)\,,\quad j=+,h
\nonumber\\
 & & \hspace{0.0cm}\chi_+=\left(\begin{array}{c}1 \\ 0\end{array}\right),\,\,
\chi_h=\left(\begin{array}{c}0 \\ 1\end{array}\right)\,.\label{tau_spins}
\end{eqnarray}
It is easy to verify that (\ref{H_tb}) in the sector $N_+=Q$ can be
mapped on the $XX$ spin model in two dimensions:
\begin{eqnarray}
H_t & = & -\frac{1}{2}\sum_{\langle x,y\rangle}\Big(\tau_1(x)\tau_1(y)+
\tau_2(x)\tau_2(y)\Big)\,.\label{H_xx}
\end{eqnarray}
The Hamiltonian (\ref{H_xx}) is invariant under the inversion of all
``$\tau$ spins'' (\ref{tau_spins})
\begin{eqnarray}
UH_tU^+\,\,=\,\,H_t & \quad & U=\prod_x\tau_1(x)
\end{eqnarray}
which explains the hole--particle symmetry (\ref{E_b_ph}).

Note, that all the matrix elements in $H_t$ have a negative sign. Therefore
we can apply the Perron-Frobenius theorem,\cite{footnote3}
which states that the ground
state $\psi_F(S_z=Q/2)$ is nondegenerate and that in the defining basis
all components are greater than zero.

Since $H_t$ conserves the total spin (\ref{s_total}), the application
of lowering operators
\begin{eqnarray}
S_- & = & \frac{1}{2}\sum_{\langle x,y\rangle}\Big(\sigma_1(x)-i\sigma_2(y)\Big)
\end{eqnarray}
onto the ground state $\psi_F(S_z=Q/2)$ with eigenvalue $E_F(Q)$ yields
eigenstates of $H_t$:
\begin{eqnarray}
(S_-)^n\psi_F(S_z=Q/2) & = & \langle\psi(Q/2)|S_+^nS_-^n|\psi(Q/2)\rangle^{-1/2}
\times\nonumber\\
 & & \psi_F(S_z=Q/2-n)\nonumber\\ & & \\[10pt]
H_t\psi_F(S_z=Q/2-n) & = & E_F(Q)\psi_F(S_z=Q/2-n)\nonumber\\
\end{eqnarray}
in the sector with $S_z=Q/2-n$ with the same eigenvalue $E_F(Q)$. Indeed these
states are again ground states in the sector $S_z=Q/2-n$ for the following
two reasons:

\begin{itemize}
\item[i)]
The state $\psi_F(S_z=Q/2-n)$ has again only positive components in the
defining basis in the sector $S_z=Q/2-n$.

\item[ii)]
The matrix elements of $H_t$ in the sector $S_z=Q/2-n$ are all negative, such 
that the Perron-Frobenius theorem can be applied again: The ground state is
nondegenerate and has only positive components. Therefore $\psi_F(S_z=Q/2-n)$
must be the ground state for $S_z=Q/2-n$.

\end{itemize}

Finally, we want to stress that all the states $\psi_F(S_z=Q/2-n)$ are
symmetric under all permutations:
\begin{eqnarray}
\Big(P(x,y)-1\Big)\Big(n_+(x)n_-(y)+n_-(x)n_+(y)\Big)\times & & \nonumber\\
\psi_F(S_z=Q/2-n) & = & 0\nonumber\\
\end{eqnarray}
for the quantum numbers of the particles at sites $x$ and $y$, which
guarantees that
\begin{eqnarray}
\vec S^2\psi_F(S_z=Q/2-n) & = & S(S+1)\psi_F(S_z=Q/2-n)\nonumber\\
 & & S=Q/2\,.\nonumber\\
\end{eqnarray}

\end{appendix}







\begin{thebibliography}{90}
\bibitem{footnote1}It should be noted, that a bosonic Hubbard model
leads to a ferromagnetic coupling in the limit $U\gg t$.
Such a model will be investigated elsewhere.
\bibitem{footnote2}Except for the incorrect ground state energy for the case
$N=4\times 4$, $Q=11$, where the true ground state [$E_{0,f}(Q=11,\alpha=0)=-12.1808457$]
has total spin $S=1/2$
-- as previously noted by Fehske\cite{fehske91} in his Habilitation
thesis.
\bibitem{footnote3}The required connectivity
condition is given for the considered lattices.
\end{thebibliography}
\end{document}